\documentclass[12pt,draftcls, onecolumn]{IEEEtran}

\usepackage{graphicx}
\usepackage{amsmath}
\usepackage{amsfonts}
\usepackage[nomarkers,nofiglist,notablist]{endfloat}
\let\MYoriglatexcaption\caption
\renewcommand{\caption}[2][\relax]{\MYoriglatexcaption[#2]{#2}}

\renewcommand{\eqref}[1]{(\ref{#1})}

\begin{document}

\title{Protocol Coding through Reordering of User Resources, Part II: Practical Coding Strategies}
\author{\begin{tabular}{c}
Petar~Popovski$^*$, Zoran~Utkovski$^{\dagger}$ and K.~F.~Trillingsgaard$^*$\\
$*$ Department of Electronic Systems, Aalborg University, Denmark \\
$\dagger$ Institute of Information Technology, University of Ulm, Germany \\
Email: zoran.utkovski@uni-ulm.de, petarp@es.aau.dk,  \\
\end{tabular}}

\maketitle
\thispagestyle{empty}

\begin{abstract}
We use the term \emph{protocol coding} to denote the communication
strategies in which information is encoded through the actions
taken by a certain communication protocol. In this work we
investigate strategies for protocol coding via combinatorial
ordering of the labelled user resources (packets, channels) in an
existing, \emph{primary system}. This introduces a new,
\emph{secondary communication channel} in the existing system,
which has been considered in the prior work exclusively in a
steganographic context. Instead, we focus on the use of secondary
channel for reliable communication with newly introduced secondary
devices, that are low-complexity versions of the primary devices,
capable only to decode the robustly encoded header information in
the primary signals. In Part I of the work we have characterized
the capacity of the secondary channel through
information--theoretic analysis. In this paper we consider
practical strategies for protocol coding inspired by the
information--theoretic analysis. It turns out that the insights
from Part I are instrumental for devising superior design of
error--control codes. This is demonstrated by comparing the error
performance to the ``na\"{\i}ve'' strategy which is presumably
available without carrying out the analysis in Part I. These
results are clearly outlining both the conceptual novelty behind
the discussed concept of secondary channel as well as its
practical applicability.
\end{abstract}

\section{Introduction}
\label{sec:Introduction}

A way to introduce new features and define a new communication channel over an existing wireless system, without introducing hardware/physical layer changes, is to encode information in the actions take by the protocol of the existing (primary) communication system. We denote such a class of communication strategies by the term \emph{protocol coding}. In this work we are considering a particular type of protocol coding, in which information is encoded in the ordering of labelled resources (packets, channels) of the primary (legacy) users. For example, if in a given scheduling frame the primary system decides to send $3$ packets to Alice and $2$ packets to Bob, then the secondary transmitter gets the right to encode additional information by rearranging these packets. This can be done in $\binom{5}{2}=10$ different ways, such that in that scheduling  frame the secondary transmitter can send $\log_2 10$ bits. In this example, a scheduling frame contains $F=5$ packets and the primary decides the \emph{state} of the frame (how many packets to Alice and Bob, respectively). In each state there are different number of possible rearrangements and the problem is that the state is not controlled by the secondary, which means that the amount of information that the secondary can send it variable and unpredictable.

The restriction that stems from operation of the primary is the key feature of the communication model. In Part I~\cite{ref:Part1} we have elaborated on the capacity of the model by using a  framework where the secondary communication channel was
represented through a cascade of channels. Such a framework is alternative to a rather standard representation using Shannon's model of channels with causal channel state information at the transmitter (CSIT), but very potent in our case as we have been able to
compute the secondary capacity under quite general error model
assumptions. The capacity calculation has lead to the concept of a \emph{multisymbol}, which is a set of actual packet rearrangements that represent identical secondary input symbol.

In practice, a secondary channel can be defined over virtually any
existing wireless system and it is of interest to find the coding
strategies that are suited to a certain primary system. We can ask
the following question: If we did not have the capacity derivation
and the multisymbol framework that has emerged from it, described in ~\cite{ref:Part1}, what would
be the ``na\"{\i}ve'' way to design a code in order to communicate over
the secondary channel? Conversely, what does the information--theoretic strategy
developed in the Part I of the paper teach us about designing good
codes (signaling) strategies for this channel?

In the absence of the multisymbol framework, we can perform the
encoding in the traditional way, by taking any usual
error-correcting code and an interleaver. However, a problem
arises when we attempt to send a sequence of $F$ bits over the
secondary channel of frame length $F$, which is in a given state
$s$ which we do not control. If the sequence of bits we need to
transmit is not one of the $\binom{F}{s}$ possible symbols, we
should pick any (e. g. randomly) of the possible symbols which are
at the same (minimal) Hamming distance from the bit sequence. For
example, for a frame length $F=4$ we take four of the coded and
interleaved bits and look at the current state of the channel (how
many 1s we can transmit in the next frame). Then, we  pick any (e.
g. randomly) of the possible frames, obtained by permuting the
packets, that has minimal possible Hamming distance. For example,
when the system needs to transmit $0101$ and the state is $s=3$,
it chooses randomly between $0111$ and $1101$. However, this leads
to ambiguity since the receiver might falsely interpret the
transmitted sequence as $0111$ being or $1101$, instead of $0101$.
Hence, even when the channel does not introduce errors, there will
be decoding errors.

Therefore, we have to look at other strategies which are
applicable in the context of protocol coding. Besides its' role in
the capacity calculation, it turns out that the multisymbol
framework can be used in the construction of error-correcting
codes for secondary communication, since it gives an insight in
the coding strategies that are approaching the capacity.

The rest of the paper is organized as follows. In Section
\ref{sec:CodingErrorless} we present transmit strategies for the
errorless case, i.e. the case when the probability of error for
secondary reception is 0. In Section \ref{sec:CodingErrors} we
address the case of secondary communication when errors are
present. First we investigate a na\"{\i}ve strategy for communication,
which does not account for the specifics of secondary
communication channels. Then, we propose a coding strategy which
is inspired
from the capacity results derived 
in Part I of the paper. In Section \ref{sec:CodeDesign} we present
a trellis coding scheme where the trellis code is based on the
multisymbol framework and evaluate its' performance. Section \ref{sec:Discussion}
presents some distinguished features of
secondary channels, potential applications and provides a discussion on the   the limitations of the
presented model for protocol coding. Finally, Section \ref{sec:Conclusion} concludes the
paper and gives directions for future work.


\section{Coding for Errorless Channels}
\label{sec:CodingErrorless}

\subsection{Motivation and System Model}
In this section we introduce practical encoding strategies when the secondary channel is assumed errorless i. e. the headers (labels) of the primary packets are perfectly received at the secondary receiver. This is interesting, since using fixed-length codes necessarily leads to nonzero probability of error, despite the fact that there are no channel induced errors on the packet headers, which in turn cary the packet label, used for secondary communication. To see why this is the case, consider the example of a primary system with a scheduling frame with $F$ packets. Assume that one wants to encode secondary information by using $L$ consecutive frames. However, if the primary system decides the state to be $0$ in all frames (i.~e. all the packets are addressed to Alice), then no secondary information can be encoded, which leads to error.

We consider a simplified model by
having only two possible packet labels, i.~e. each packet in a
frame is addressed either to user $0$ or user $1$. This setup is
sufficient to illustrate the main strategies for protocol coding
with resource ordering. We recall that the set of packets that are
scheduled in a frame is decided by the primary system, i.e. the
secondary communication is restricted and can only rearrange the
set of packets selected by the primary. The state of the frame $s$
is the number of packets addressed to user $1$ and occurs with
probability $P_S(s)=\frac{\binom{F}{s}}{2^F}$. In the errorless
case, the state $s$ is known to both the transmitter and the
receiver. Thus, each state $s$ is associated with one
communication sub-channel with a capacity $r(F,s)=\log_2
\binom{F]}{s}$, as elaborated in Part I. In a given frame, the primary chooses the state $S=s$ with probability $P_S(s)$, independently of the states in the previous frames .

Let us consider $n$ frames, where each frame is a secondary channel use and
let $s^n$ be the vector that describes the random outcome from
observing the $n$ frames. As $n$ goes to infinity, the sequence
$s^n$ becomes typical, such that the number of states that will
have the value $s$ is approximately $\approx nP_S(s)$.
The capacity of the the errorless channel can be
calculated~\cite{PetarGC10}:
\begin{equation}\label{eq:CapacityComErrorless}
    C_{F,0}=\sum_{s=0}^F P_S(s) \log_2 \binom{F}{s}
\end{equation}
A scheme that achieves this capacity would work as follows.
Consider transmission of a large message by using large number of
channel uses $n \rightarrow \infty$. The sender segments the
message into sub-messages, where each of the sub-messages is sent
over a separate sub-channel (state), which occurs with probability
$P_S(s)$. The sub-message that is to be sent over the sub-channel
defined by $s$ contains approximately $nP_S(s)r_(F,s)$ bits. If
during the $i-$th channel use the sender observes that the state
$s$, then it takes the next $r(F,s)$ bits from the corresponding
sub-message. Thus, the whole message is sent by time--interleaving
of all the available sub--channels.

Nevertheless, in this paper we are focused on practical coding strategies and this scheme
might be ineffective when sending messages of finite, short length $m$. For example, when the number of channel
uses $n$ is finite, some of the channel states might not appear at
all. In this case, the above strategy would fail to send a part of
the message, i.e. the sub-messages associated with those states.
What we need is a practical coding scheme which is tailored to the
observation that the secondary system does not have a control over
the state $s$ of the channel.

\subsection{Coding with Variable Radix}

Since using a fixed-length secondary coding block introduces errors,
we need to send a group of $m$ bits by
using multiple (variable) number of frames, where the number of
frames to be used depends on the realization of the sequence of
channel states. We introduce the proposed coding strategy by
taking an example with frame of length $F=4$.

\subsubsection{Example}
Let us group the input bits into groups of size $m=10$ and let us
call such a group of bits \emph{input symbol}. Hence, one input
symbol is a number between 0 and 1023. This input symbol will be
sent by using several frames. Let the input symbol be the binary
representation of $777$, for example. Let $s_i$ be the channel
state in the $i-$th frame and let us assume the following state
sequence $s_1, s_2, s_3, s_4, s_5, s_6 \cdots= 2, 3, 0, 2, 1, 3,
\cdots$

The first state allows for $\binom{4}{2}=6$ different
combinations. We divide the interval of integers $[0-123]$
in $6$ bins such that $4$ bins have a size of $171$ and $2$ bins
have a size of $170$. The bins (sub-intervals) are given as
$[0-170]$, $[171-341]$, $[342-512]$, $[513-683]$, $[684-853]$ and
$[854-1023]$.
The number $777$ is in the bin $5$. If we enumerate the $6$
possible combinations of packets when the state is $s=2$ as $0011,
0101, 0110, 1001, 1010, 1100$, $5$ corresponds to the combination
$1010$. Hence, Alice arranges the two packets in the frame and sends
$1010$ in order to signal that the number $777$ is in the bin
number $5$. When Bob receives the first frame and sees that $s_1=2$,
then $B$ knows that it should use $6$ bins. In this way it gets
the information that the number $M$ is in the $5$-th bin, $M\in
[684-853]$.

Now Alice will use the subsequent frames in order to specify where
within the bin $[684-853]$ is the number represented by the $m=10$
bits. For the second frame, the state is $s_2=3$. The number of
possible packet rearrangements  is $\binom{4}{3}=4$. Therefore,
the bin $[684-853]$ is divided into $4$ sub-bins, such that $2$
bins have size of $43$ and $2$ bins have size of $42$. The bins
are given as $[684-726], [727-769], [770-811]$ and $[812-853]$.
Since $M=777$ belongs to the third bin, $M\in [770-811]$ and
$s=3$, in the second frame we choose the packet rearrangement
$1101$. Now, Bob knows that the transmitted number lies in the bin
$[770-811]$.

When the third frame arrives, $s_3=0$ and no information can be
sent by using combinatorial reordering.

We proceed in the same way for the rest of the frames. During the
fourth frame the channel state is $s_4=2$, which allows for $6$
different combinations, i.e the bin $[770-811]$ is divided into
$6$ sub-bins. Since $M=777$ belongs to the second bin, $M\in
[777-783]$, we send the combination (rearrangement) $0101$. During
the fifth frame the channel state is $s_5=1$, which allows for $4$
different combinations, i.e there are $4$ sub-bins. Since $M=777$
belongs to the first bin, $M\in [777-783]$, we send the
combination (rearrangement) $0001$. During the sixth frame the
channel state is $s_6=3$, which allows for $4$ different
combinations. We divide the bin $[777-783]$ in $4$ sub-bins, given
as $[777-778], [779-780], [781-782], [783]$. Since $M=777$ belongs
to the first bin, $M\in [777-783]$, we send the combination
(rearrangement) $0001$. Bob sees that the channel state is $s_5=1$,
and knows that there are $4$ bins and that $M\in [777-778]$. This
is over-dimensioned for the two bins we have left. Nevertheless, Alice
can send $0011$ in order to inform Bob that the number is $M=777$.

Alternatively, Alice can apply additional optimization. Since $s_6=3$,
the representation of the numbers in the bin $[777, 778]$ is
overprovisioned.  Since the fourth frame can have $4$ possible
combinations, it means that it can represent the $2$ possible
numbers in the bin and also $2$ possible bins for the next input
symbol. Let the next input symbol be $233$. We divide the interval
$[0,1023]$ in $4$ bins, such that the $4$ bins have a size of
$256$. Hence, the number $233$ is in bin $1$. This bin can be sent
jointly with the last bin of the first input symbol. Now the
interval $[1-4]$ is divided into $2$ bins of size $1$: $[1-2],
[3-4]$. A needs to use the first bin, since $777$ is the first
number in the uncertainty window $[777, 778]$ for the first input
symbol, but it will send the combination number $3$ (0110) in
order to inform that $233$ is in the second bin of $1024$ when
$1024$ is divided in $4$ bins as described above.

We note that for a different channel state sequence, the number of
frames over which the coding was performed would differ. Hence, we
speak of a \emph{variable-radix} scheme which uses a variable number of
frames to represent the secondary information bits. The variable
radix scheme provides an effective mapping of the information bits
into symbols, taking into account that the communication takes
place over a multiple-state channel. Indeed, without having
control of the channel state $s$, we can not send the sequence of
information bits by using a single frame of length $F$ as the
number of different combinations $\binom{F}{s}$ may not be
sufficient. On the other hand, by taking multiple frames and
allowing the number of frames to vary, this is always possible, as
shown in the example. In this sense we can think of this scheme as
an efficient modulation scheme, since it primarily provides
mapping of the information bits into symbols, by taking into
account the unpredictable channel states.

The example described above should be sufficient for the reader to devise variable-radix schemes for another state sequence.
The problem with the variable-radix scheme is that it works only
under perfect, errorless conditions and in the next section we discuss the practical strategies for protocol coding that can deal with channel -induced errors.

\section{Coding for Channels with Errors}
\label{sec:CodingErrors}
As already argued, the variable-radix
scheme fails in the presence of channel errors and might lead to
catastrophic behavior. For example, if the channel is mostly
errorless, but one symbol is in error, then the receiver Bob will not know where
the current symbol ends and where the next starts, such that the whole sequence of information bits will be in error.

With this argument on mind, in the case with channel errors we
would prefer to code over a fixed number of frames (and thus allow errors even when there are no channel-induced errors on the packet labels). The question
is which coding strategy would be applicable in this case. We
first look at a na\"{\i}ve coding strategy which does not account for
the specifics of the secondary channel. Later, we present a
communication scheme which is inspired by the capacity results
derived previously

\subsection{Na\"{\i}ve Coding Strategy}
The na\"{\i}ve strategy works as follows. We take any usual
error-correction code of rate $R$ and interleave the output of this
code, e. ~g. by using a pseudo-random interleaver. The motivation for using an interleaver is to break the burst bit errors that can occur within one secondary symbol (frame). For example, for
a frame length $F=4$ we take four of the coded and interleaved
bits and look at the current state of the channel (how many 1s we
can transmit in the next frame). Then, we  pick any (e. g.
randomly) of the possible frames, obtained by permuting the
packets, that has minimal possible Hamming distance. For example, let the coded bits are 0101 and let the state be
$s=3$. Then the Hamming distance of the ``true information'' 0101
from 0111, 1101 is 1 (minimal possible), while it is 3 from 1011
or 1110. Hence, when the system needs to transmit 0101 and the
state is $s=3$, it chooses randomly between 0111 and 1101.

The trouble with the described na\"{\i}ve strategy is that, even when
the channel does not introduce errors, there will be decoding
errors. Additionally, as we will see in Section
\ref{sec:CodeDesign}, the na\"{\i}ve strategy has poor performance in
channels with transmission errors.

\subsection{Coding Strategy inspired from the Information--Theoretic Analysis}
\label{sec:ITCoding} Here we propose a coding strategy which is
inspired by the capacity results for the secondary communication
channel derived in Part I. We start by recalling that we modelled
the secondary communication channel by using a cascade of two
channels
$T - \mathbf{X} - \mathbf{Y}$ 
where the input constraints from the primary system were reflected
in the way the set ${\cal P}_{\mathbf{X}|T}$ is defined. Then,
instead of speaking about which strategies out of ${\cal T}$ that
are chosen with non--zero probability, we speak of which
representatives to choose for given secondary input symbol $T=t$. According to this, we
introduced the notion of multisymbol, which is the set of
representatives $\mathcal{M}_t=\{ \mathbf{x}_s(t) \}$ for given
$t$. Further, a minimal multisymbol was defined as a multisymbol
obtained by permutation of the basic multisymbol, where the basic
multisymbol
has 
representatives defined as follows:
\begin{equation}
x_s(t)=\left\{
\begin{array}{rl}
0 & \text{if } i \leq F-s \\
1 & \text{otherwise } \end{array} \right.\nonumber
\end{equation}
The minimal multisymbols satisfy the following Hamming distance relation $d_H(\mathbf{x}_s(t),\mathbf{x}_{s+1}(t))=1$
for each $s=0 \ldots F-1$. The main capacity result for the
secondary channel obtained in Part I has been stated in the
following form
\begin{equation}
   C \leq C_{XY}- I_m.\nonumber
\end{equation}
The term $C_{XY}$ is the capacity of the $\mathbf{X}-\mathbf{Y}$
channel, given by (12) in Part I. We recall that this capacity is
attained by the distribution
$\sum_{\mathbf{x} \in {\cal X}_s} P_{\mathbf{X}}(\mathbf{x}) =
P_S(s)$
and is upper bounded by the capacity of the channel defined by the
underlying error model. $I_m$ is the minimal value (constant) of
$I(\mathbf{X};\mathbf{Y}|T=t)$, achieved by the choice of the
multisymbol as a minimal multisymbol. The equality is achieved if
and only if there is a pair of distributions $\left ( P_T(\cdot),
P_{\mathbf{X}|T}(\cdot) \right )$ that simultaneously attains the
maximum and the minimum in the first and the second term,
respectively.

This result is quite general and holds for all classes of
memoryless channels $\mathbf{X}-\mathbf{Y}$ with binary inputs.
Among other channels, it holds for the erasure channel, the binary
symmetric channel and the $Z$ channel. It has been further shown
that for a uniform distribution over ${\cal T}$, this capacity can
be achieved by a set $\mathcal{T}$ of cardinality
$L=\textrm{lcm}\left(\binom{F}{0},\binom{F}{1},\ldots,\binom{F}{F}\right)$.
The $L$ multisymbols should be minimal, meaning that the Hamming
distance between two adjacent symbols is 1, $d_H(x_s,x_{s+1})=1$.

From the viewpoint of capacity, the choice of the multisymbols is
irrelevant, as long they are minimal and the distribution of $X$
fulfills the required condition. However, the choice of the
multisymbols does affect the performance of the error-correcting
code constructed based on the multisymbol framework.

Our aim is to use the multisymbol framework in the construction of
practical coding schemes which are better suited for the secondary
communication channel than the na\"{\i}ve approach. The question
to ask is which criterion, e.g. distance metric we are going to
use in the selection of the multisymbols. We adopt a heuristic
approach and take the \textit{expected Hamming distance} as the
metric of interest for the choice of the $L$ multisymbols. The
expected Hamming distance for two multisymbols $t_1$ and $t_2$ is
defined as follows
\begin{equation}
E_{d_H}(t_1,t_2)=\sum_{s=0}^{F} P_S(s) d_H(x_s(t_1), x_s(t_2)),
\end{equation}
where $d_H$ is the Hamming distance between the two vectors.
Clearly, considering the triviality of the states $s=0$ and $s=F$,
we can simplify to:
\begin{equation}
E_{d_H}(t_1,t_2)=\sum_{s=1}^{F-1} P_S(s) d_H(x_s(t_1), x_s(t_2))
\end{equation}
The motivation behind this is that this metric incorporates the
state of the channel which can not be controlled by the secondary
system.

With this in mind, we can construct a convolutional code by using
the multisymbols framework and the expected Hamming distance as
design criterion. We define a trellis for the convolutional code
with a certain number of states. In the trellis diagram, each
state contains two outgoing paths, each of them corresponding to
one possible input binary symbol. Also, each state has two
incoming paths. Each branch in the trellis is associated with an
input symbol and an output symbol. In our case, the input symbol
is binary and the output symbol is one of the $L$ multisymbols.
The trellis is chosen on purpose to have $L$ branches, such that
each multisymbol is used only once.

Now, the question is how we associate multisymbols with the
transitions in the trellis. We use the known rules from trellis coding:
the output symbols on the branches exiting from the same state
should be maximally separated in terms of the expected Hamming
distance. The same is valid for the output symbols associated with
the two branches that enter the same state. In order to illustrate the code construction,
we take the example with $F=4$, where the minimal cardinality of
the uniform auxiliary variable $T$ is
$L=\mathrm{lcm}\{\binom{4}{0},\binom{4}{1},\ldots,\binom{4}{4}\}=12$.

There are multiple ways in which the multisymbols can be chosen,
and different sets have different features. We can get useful
insights about the expected Hamming distance spectrum if we use
the representation of the multisymbols as paths in the directed
graph, as shown in Fig.~\ref{fig:Set1}. In order to maximize the
expected Hamming distance between multisymbols, the paths
corresponding to the multisymbols should be as diverse as
possible. To assure this, we have to choose the multisymbols such
to avoid (as much as possible) having multisymbols with common
edges. Indeed, for two different multisymbols $t_1$ and $t_2$
which share a common edge $(x_j(t),x_{j+1}(t))$, the terms in the
expected Hamming distance
\begin{equation}
E_{d_H}(t_1,t_2)=\sum_{s=0}^{F} P(S=s) d_H(x_s(t_1), x_s(t_2)),
\end{equation}
associated with that edge will be $0$. The necessary condition to avoid a common edge
between the nodes from $\mathcal{X}_s$ and $\mathcal{X}_{s+1}$,
where $s\leq \lfloor F/2 \rfloor -1$, is that $L/\binom{F}{s}\leq
F-s$. In other words, the edge weight should be at most $1$.

We note once again that the relevance of the expected Hamming
distance as a performance metric is stated as a conjecture which
is not rigorously proved. Moreover, one has to look at the whole
distance spectrum, in order to be able to predict the performance
of the error-correcting code. Since in the general case it is
difficult to control the code distance spectrum, we turn to the
the minimal expected Hamming distance as a simplified indicator of
the code performance. However, in this particular case we have to
be careful when making conclusions about the performance which are
based solely on the minimal distance. As we are going to see in
the next section, some of the simulation results indicate that the
minimal expected Hamming distance is not the only factor which is
decisive for the error performance.

In the following we give three representative examples of the set
of multisymbols $\mathcal{T}$, created for $F=4$.

\subsubsection{Choice of multisymbols, Example 1}
\begin{figure}

\begin{minipage}[b]{0.5\linewidth}
\centering { \footnotesize
\begin{tabular}{|c|c|}
\hline
$t$ & $ \{ \mathbf{x}_s(t) \}$ \\
\hline
$1$ & $(0000,0001,0011,0111,1111)$\\
\hline
$2$ & $(0000,0001,0101,1101,1111)$ \\
\hline
$3$ & $(0000,0001,1001,1011,1111)$ \\
\hline
$4$ & $(0000,0010,0011,0111,1111)$ \\
\hline
$5$ & $(0000,0010,1010,1011,1111)$ \\
\hline
$6$ & $(0000,0010,1010,1110,1111)$ \\
\hline
$7$ & $(0000,0100,0101,1101,1111)$ \\
\hline
$8$ & $(0000,0100,0110,0111,1111)$ \\
\hline
$9$ & $(0000,0100,0110,1110,1111)$ \\
\hline
$10$ & $(0000,1000,1001,1011,1111)$ \\
\hline
$11$ & $(0000,1000,1100,1101,1111)$ \\
\hline
$12$ & $(0000,1000,1100,1110,1111)$ \\
\hline
\end{tabular}
}

{\footnotesize (a)}
\end{minipage}
\hspace{0.5cm}
\begin{minipage}[b]{0.5\linewidth}
\centering
\includegraphics[width=8cm]{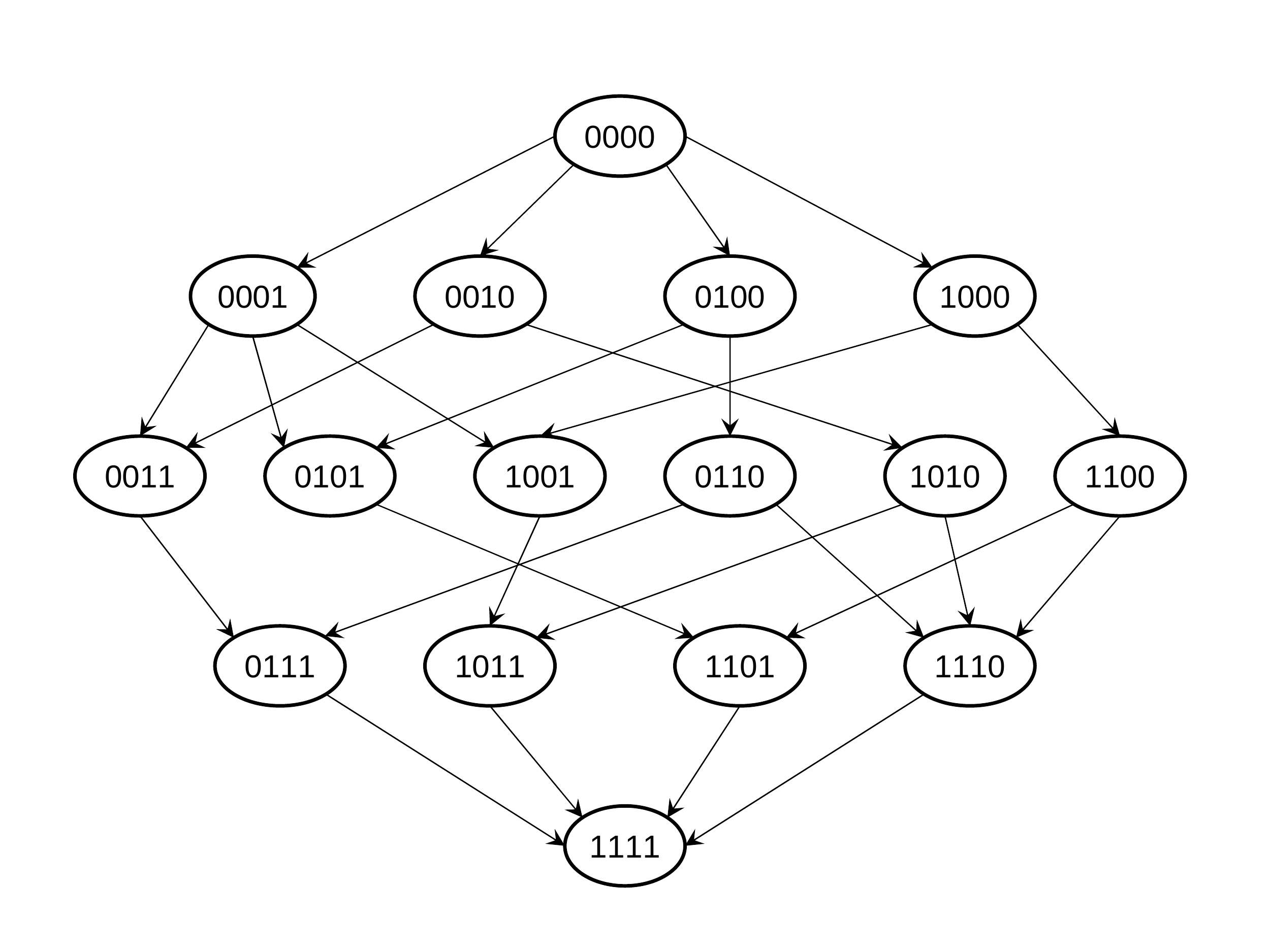}
{\footnotesize (b)} 
\end{minipage}

\caption{Selection of the representative sets for $F=4$, Example
1. The selection of the multisymbols is not optimal. (a)
Multisymbols for the $12$ inputs. (b) Graph representation of the
process for selecting the multisymbols $\mathbf{x}_s(t)$.}
\label{fig:Set1}
\end{figure}

In the first example, we choose the $12$ multisymbols as given in
Fig.~\ref{fig:Set1} a). The multisymbols are chosen as a
permutation of the basic multisymbol
$\mathcal{M}_b=\{0000,0001,0011,0111,1111\}$ and fulfill the
required property about the distribution of $X$. Hence, this
choice is capacity achieving, but we need to investigate its
performance in terms of error rate when used to construct a
channel code. We use the representation of the multisymbols as
paths in the directed graph, as shown in Fig.~\ref{fig:Set1} b).
We note that in the graph representation, some of the multisymbols
have common edges which can be avoided. This, for example, is the
case with the multisymbols $t_1=\{0000,0001,0011,0111,1111\}$ and
$t_4=\{0000,0010,0011,0111,1111\}$.  The expected Hamming distance
profile for the above choice of the set of multisymbols reveals
that the minimal distance is $0.5$.

\subsubsection{Choice of multisymbols, Example 2}
\begin{figure}
\begin{minipage}[b]{0.5\linewidth}
\centering { \footnotesize
\begin{tabular}{|c|c|}
\hline
$t$ & $ \{ \mathbf{x}_s(t) \}$ \\
\hline
$1$ & $(0000,0001,0011,0111,1111)$\\
\hline
$2$ & $(0000,0001,0101,1101,1111)$ \\
\hline
$3$ & $(0000,0001,1001,1011,1111)$ \\
\hline
$4$ & $(0000,0010,0011,1011,1111)$ \\
\hline
$5$ & $(0000,0010,0110,0111,1111)$ \\
\hline
$6$ & $(0000,0010,1010,1110,1111)$ \\
\hline
$7$ & $(0000,0100,0101,0111,1111)$ \\
\hline
$8$ & $(0000,0100,0110,1110,1111)$ \\
\hline
$9$ & $(0000,0100,1100,1101,1111)$ \\
\hline
$10$ & $(0000,1000,1001,1101,1111)$ \\
\hline
$11$ & $(0000,1000,1010,1011,1111)$ \\
\hline
$12$ & $(0000,1000,1100,1110,1111)$ \\
\hline
\end{tabular}
}

{\footnotesize (a)} 
\end{minipage}
\hspace{0.5cm}
\begin{minipage}[b]{0.5\linewidth}
\centering { \footnotesize
\begin{tabular}{|c|c|}
\hline
$t$ & $ \{ \mathbf{x}_s(t) \}$ \\
\hline
$1$ & $(0000,0001,0011,0111,1111)$\\
\hline
$2$ & $(0000,0001,0101,0111,1111)$ \\
\hline
$3$ & $(0000,0001,1001,1011,1111)$ \\
\hline
$4$ & $(0000,0010,0011,1011,1111)$ \\
\hline
$5$ & $(0000,0010,0110,0111,1111)$ \\
\hline
$6$ & $(0000,0010,1010,1011,1111)$ \\
\hline
$7$ & $(0000,0100,0101,1101,1111)$ \\
\hline
$8$ & $(0000,0100,0110,1110,1111)$ \\
\hline
$9$ & $(0000,0100,1100,1101,1111)$ \\
\hline
$10$ & $(0000,1000,1001,1101,1111)$ \\
\hline
$11$ & $(0000,1000,1010,1110,1111)$ \\
\hline
$12$ & $(0000,1000,1100,1110,1111)$ \\
\hline
\end{tabular}
}

{\footnotesize (b)} 
\end{minipage}\\

\vspace{1cm}

\begin{minipage}[b]{\linewidth}
\centering
\includegraphics[width=8cm]{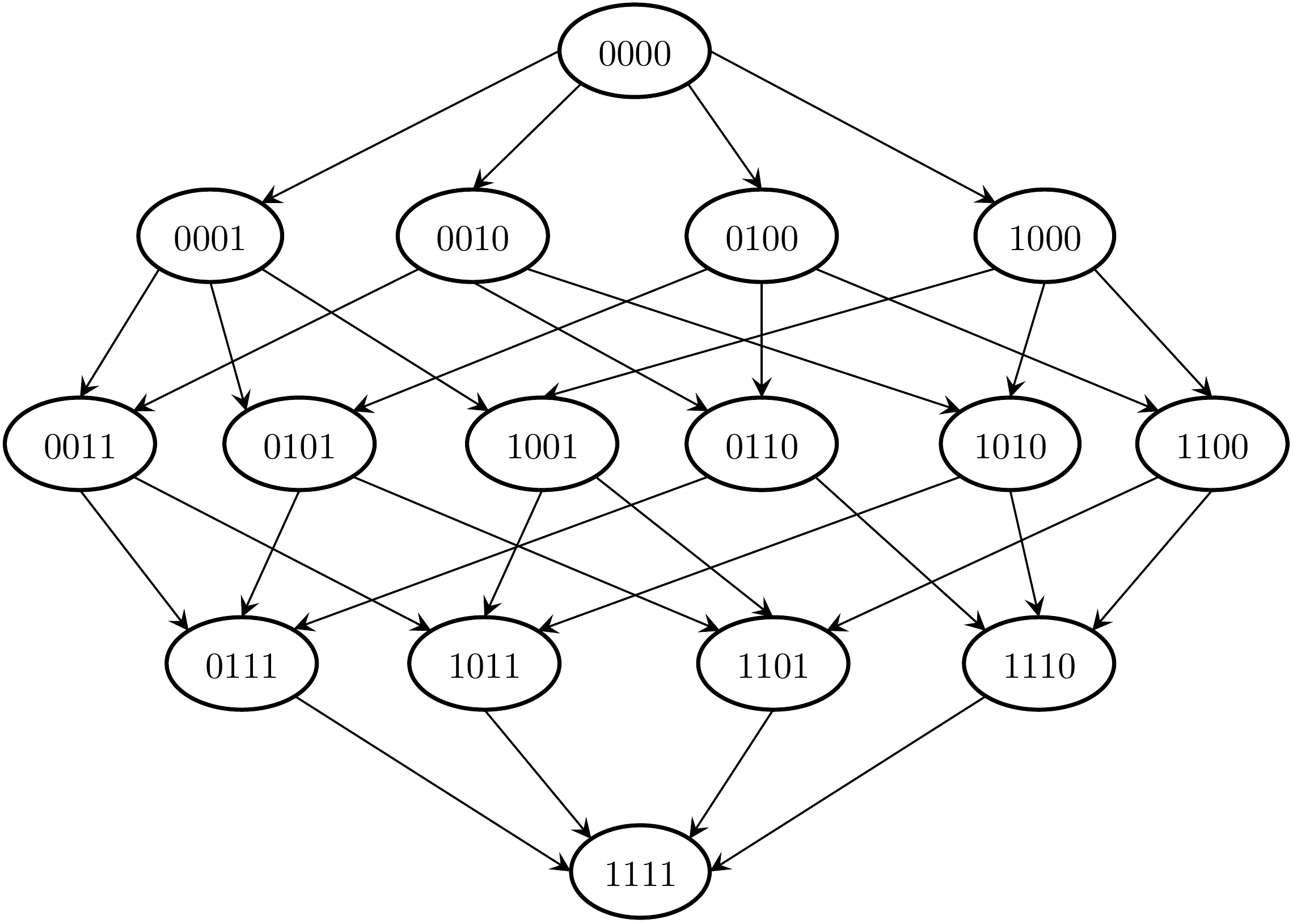}\\
{\footnotesize (c)} 
\end{minipage}
\caption{Selection of the representative sets for $F=4$ that
achieve the capacity, Example 2. (a) Multisymbols for the $12$
inputs with minimal expected Hamming distance $1$. (b)
Multisymbols for the $12$ inputs with minimal expected Hamming
distance 0.75. (c) Graph representation of the process for
selecting the multisymbols $\mathbf{x}_s(t)$. Both sets are
constructed based on the same graph representation, only the
selected paths in the graph for the corresponding multisymbols are
different in the two sets.} \label{fig:Set2}
\end{figure}

In the second example, we choose the $12$ multisymbols as given in
Fig.~\ref{fig:Set2} a). We observe that no two multisymbols are
identical and the choice of the multisymbols is capacity
achieving. The graph used for selection of the multisymbols is
shown in Fig.~\ref{fig:Set2} c). The multisymbols are constructed
by using each edge of the graph exactly once, except for the edges
between $\mathcal{X}_0=\{0000\}$ and
$\mathcal{X}_1=\{0001,0010,0100,1000\}$, where common edges can
not be avoided. Additionally, common edges are avoided later in
the graph, by an adequate choice of the paths associated with the
multisymbols. For example, we choose
$t_2=\{0000,0001,0101,1101,1111\}$ instead of
$t_2=\{0000,0001,0101,0111,1111\}$ in order to avoid a common edge
with $t_1=\{0000,0001,0011,0111,1111\}$ in the last section of the
graph. The minimal expected Hamming distance for this choice of
multisymbols is $1$. We expect that this set will perform better
compared to the set of multisymbols in Example 1, due to the
better distance spectrum. This conjecture is confirmed in Section
\ref{sec:CodeDesign} where we present the simulation results for
the performance of these choices of multisymbols.


The previous observations lead us to ask if there is a general
strategy that produces the choice of the set of multisymbols for
an arbitrary $F$ which has the maximal minimal expected Hamming
distance? The answer is affirmative. Without giving a detailed
proof, we will only note that this set is implicitly constructed
in Appendix, Part I~\cite{ref:Part1}, where it is shown that it is
always possible to choose the
$L=\mathrm{lcm}\{\binom{F}{0},\binom{F}{1},\ldots,\binom{F}{F}\}$
paths in the direct graph. A careful examination of the
argumentation in the proof reveals that a set of $L$ multisymbols
with the required properties can be constructed by the procedure
described in the proof.

Surprisingly, we have been able to find a set of multisimbols with
minimal expected Hamming distance $0.75$ which performs better
than the above set with minimal distance $1$, as presented in the
simulation results in the next section. The set of multisymbols is
presented on Fig.~\ref{fig:Set2} b) and is obtained by using the
same graph representation Fig.~\ref{fig:Set2} c), only using
different paths in the graph. We suspect that the reason for this
behavior is that the minimal distance itself is not decisive for
the performance, even if the conjecture that the expected Hamming
distance is the relevant metric for the error-control coding
holds. Probably, one has to look at both the distance spectrum and
the trellis diagram in details, in order to make the right
conclusion about the code performance. Nevertheless, the
performance in both cases is superior to the na\"{\i}ve scheme,
which makes the case for the relevance of the multisymbol
framework in the design of practical error-control schemes.

\subsubsection{Choice of the multisymbols, Example 3}
\begin{figure}
\begin{minipage}[b]{0.5\linewidth}
\centering { \footnotesize
\begin{tabular}{|c|c|}
\hline
$t$ & $ \{ \mathbf{x}_s(t) \}$ \\
\hline
$1$ & $(0000,0001,0011,0111,1111)$\\
\hline
$2$ & $(0000,0001,0011,0111,1111)$ \\
\hline
$3$ & $(0000,0001,0101,0111,1111)$ \\
\hline
$4$ & $(0000,0010,0110,1110,1111)$ \\
\hline
$5$ & $(0000,0010,1010,1011,1111)$ \\
\hline
$6$ & $(0000,0010,1010,1011,1111)$ \\
\hline
$7$ & $(0000,0100,0101,1101,1111)$ \\
\hline
$8$ & $(0000,0100,0110,1110,1111)$ \\
\hline
$9$ & $(0000,0100,1100,1101,1111)$ \\
\hline
$10$ & $(0000,1000,1001,1101,1111)$ \\
\hline
$11$ & $(0000,1000,1001,1101,1111)$ \\
\hline
$12$ & $(0000,1000,1100,1110,1111)$ \\
\hline
\end{tabular}
}

{\footnotesize (a)} 
\end{minipage}
\hspace{0.5cm}
\begin{minipage}[b]{0.5\linewidth}
\centering
\includegraphics[width=8cm]{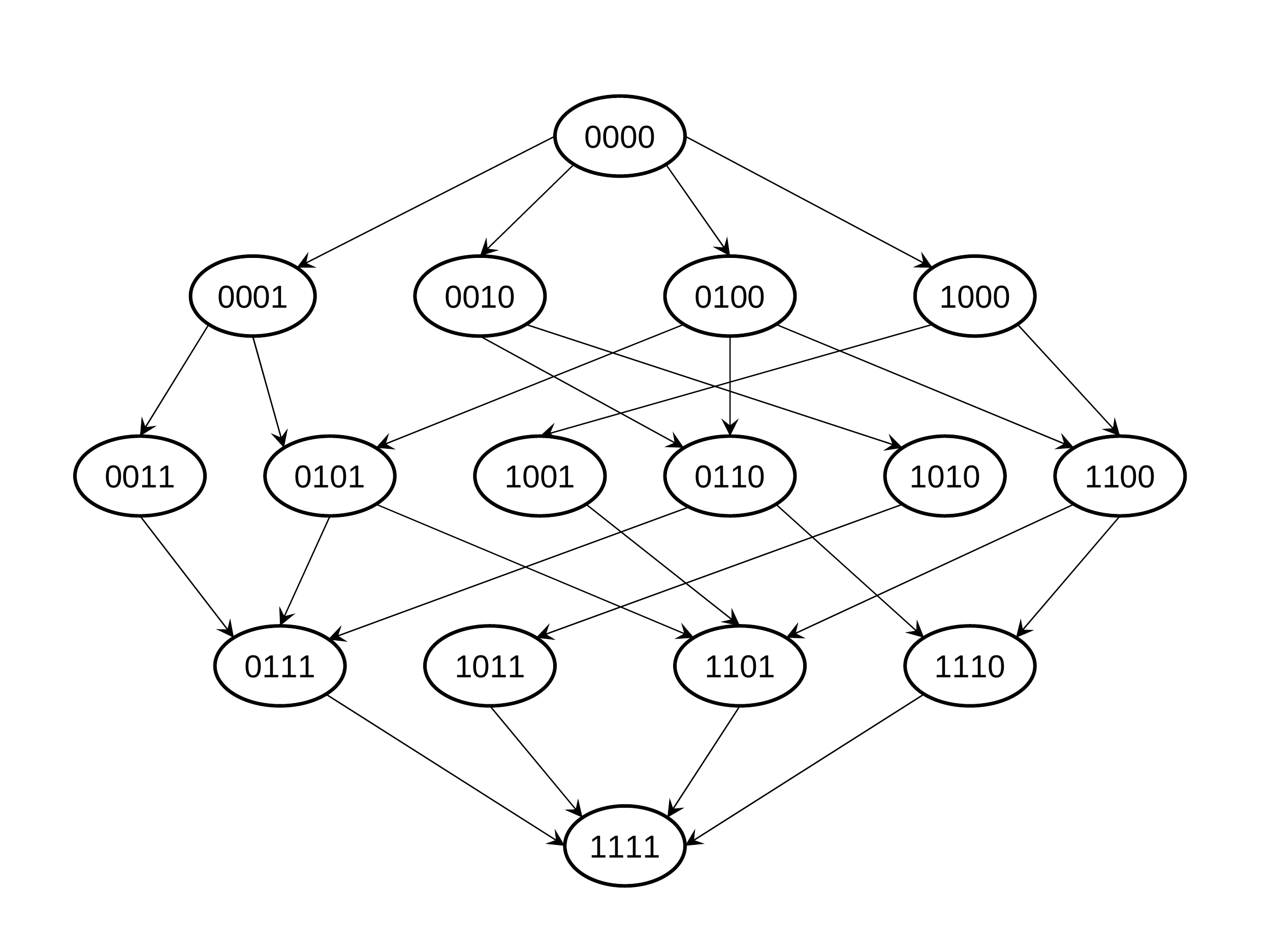}\\
{\footnotesize (b)} 
\end{minipage}
\caption{Selection of the representative sets for $F=4$, Example
3. The selection of the multisymbols yields non-uniform
distribution of $\mathcal{T}$. (a) Multisymbols for the $12$
inputs (b) Graph representation of the process for selecting the
multisymbols $\mathbf{x}_s(t)$.} \label{fig:Set3}
\end{figure}

As a third example, we choose the $12$ multisymbols as shown in
Fig.~\ref{fig:Set3} a). We notice that, according to this choice,
the multisymbols $t_1$ and $t_2$ are identical. Actually, besides
$t_1=t_2$, we have also $t_5=t_6$ and $t_{10}=t_{11}$. We note
that this choice does not violate the conditions for minimal
multisymbols and satisfies the target distribution over $X$, thus
it is capacity achieving.
At the first sight, this result seems counterintuitive and reveals the following problem: why we do not lose capacity even if we are not
using the highest possible diversification at the input (in this
case we assign the same multisymbol to two input symbols $t$)?
We note that the cardinality of the input
symbols is not $12$, but $9$. However, they are non-uniformly
distributed --- for example, $6$ have probability $\frac{1}{12}$ and $3$ have probability
$\frac{1}{6}$. Until now, we have constrained ourselves to uniform
distribution over the input symbols. However, it can be shown that
if non--uniform distribution is used over ${\cal T}$, then
capacity can be achieved even with $|{\cal
T}|<L=\textrm{lcm}\left(\binom{F}{0},\binom{F}{1},\ldots,\binom{F}{F}\right)$.
For this particular instance with $F=4$, it can be shown that in the above example the capacity
can be achieved by a set $\mathcal{T}$ with cardinality $|{\cal
T}|=8$. The probability distribution of the input symbols ${\cal
T}$ is $P_T(t)=\frac{1}{6}$ for $t=1, 2, 3, 4$ and
$P_T(t)=\frac{1}{12}$ for $t=5, 6, 7, 8$. In the following we
specify only the nonzero members  $\mathbf{P}$, the transition
matrix for the channel $T-\mathbf{X}$. Note that the notation is
slightly abused, with e.~g. $P(0001|T=1,5)$ meaning $P(0001|T=1)$
or $P(0001|T=5)$): $P(0000|T=t) = P(1111|T=t)=\frac{1}{16} \quad
\textrm{for any }t=1 \ldots 8$; $P(0001|T=1,5) = P(0010|T=2,6) =
P(0100|T=3,7)=P(1000|T=4,8)=\frac{4}{16}$ $P(0011|T=1)
=P(0110|T=2)=P(1100|T=3)=P(1001|T=4)=\frac{6}{16}$; $P(0101|T=5,7)
=P(1010|T=6,8)=\frac{6}{16}$, and $P(0111|T=1,5) = P(1110|T=2,6) =
P(1101|T=3,7)=P(1011|T=4,8)=\frac{4}{16}$. The general case of $T-\mathbf{X}$ with non--uniform distribution
on ${\cal T}$ and minimal required size $|{\cal T}|$ to achieve
the capacity is outside of the scope for this paper and is a topic
of ongoing work.

In the following section we present worked-out examples of trellis
codes based on the multisymbol framework.

\section{Code Design and Simulation Results}
\label{sec:CodeDesign}

\subsection{Code Design}
\begin{figure}
\begin{minipage}[b]{\linewidth}
\centering
\includegraphics[width=12cm]{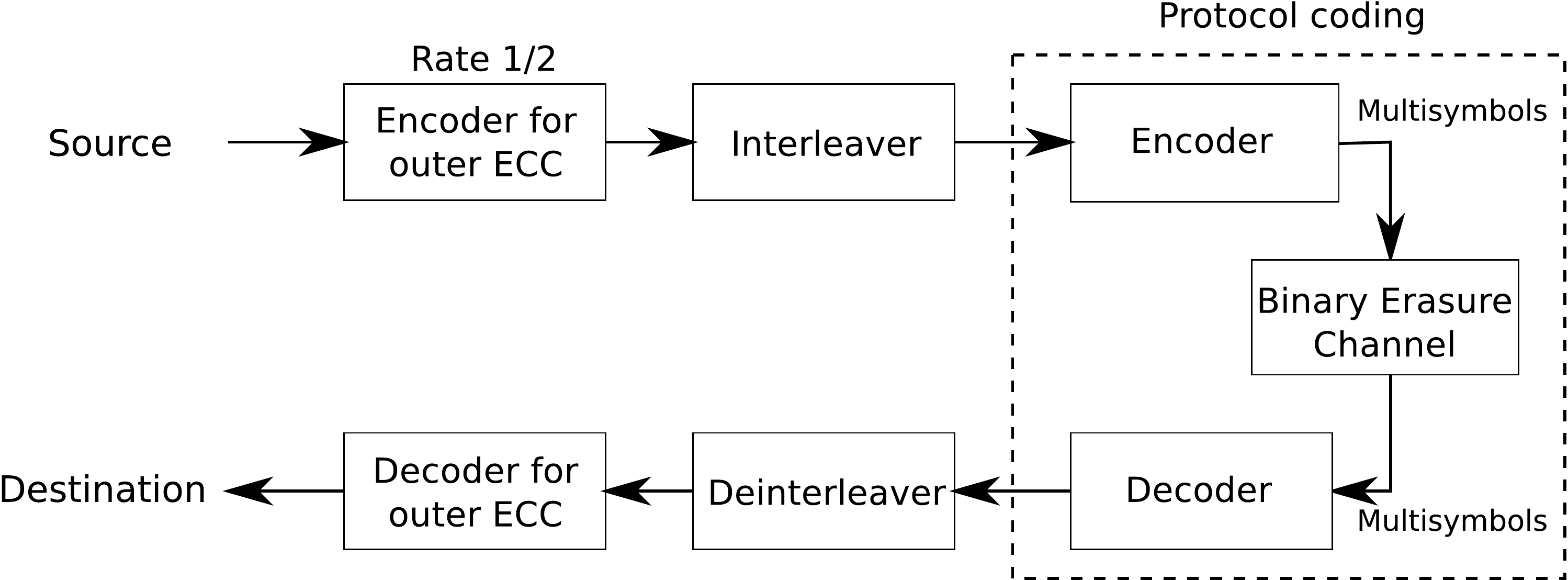}\\
{\footnotesize (a)}\\
\end{minipage}

\vspace{2cm}
\begin{minipage}[b]{0.5\linewidth}

\centering
\includegraphics[width=5cm]{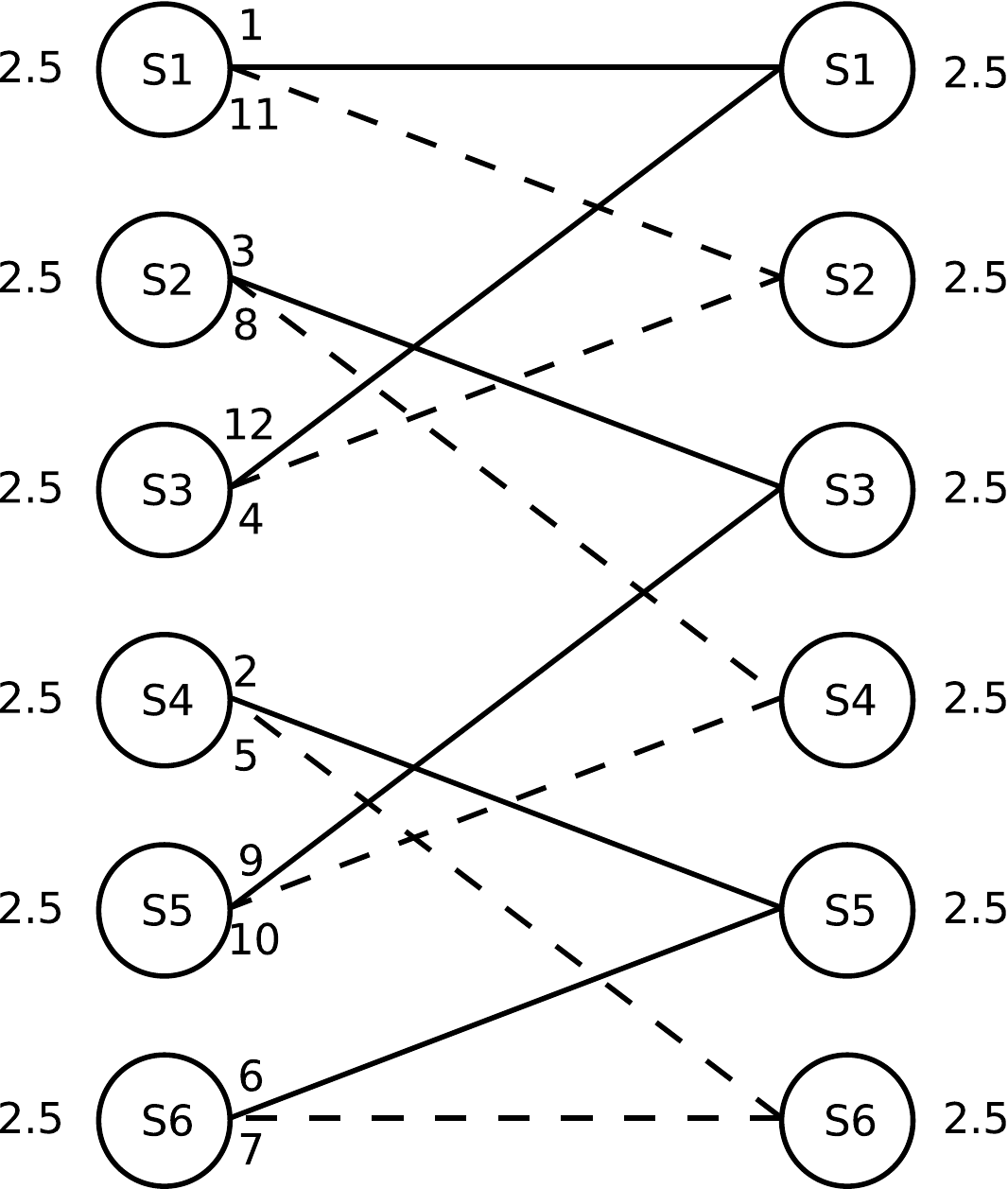}\\
{\footnotesize (b)} 
\end{minipage}
\hspace{0.5cm}
\begin{minipage}[b]{0.5\linewidth}
\centering
\includegraphics[width=5cm]{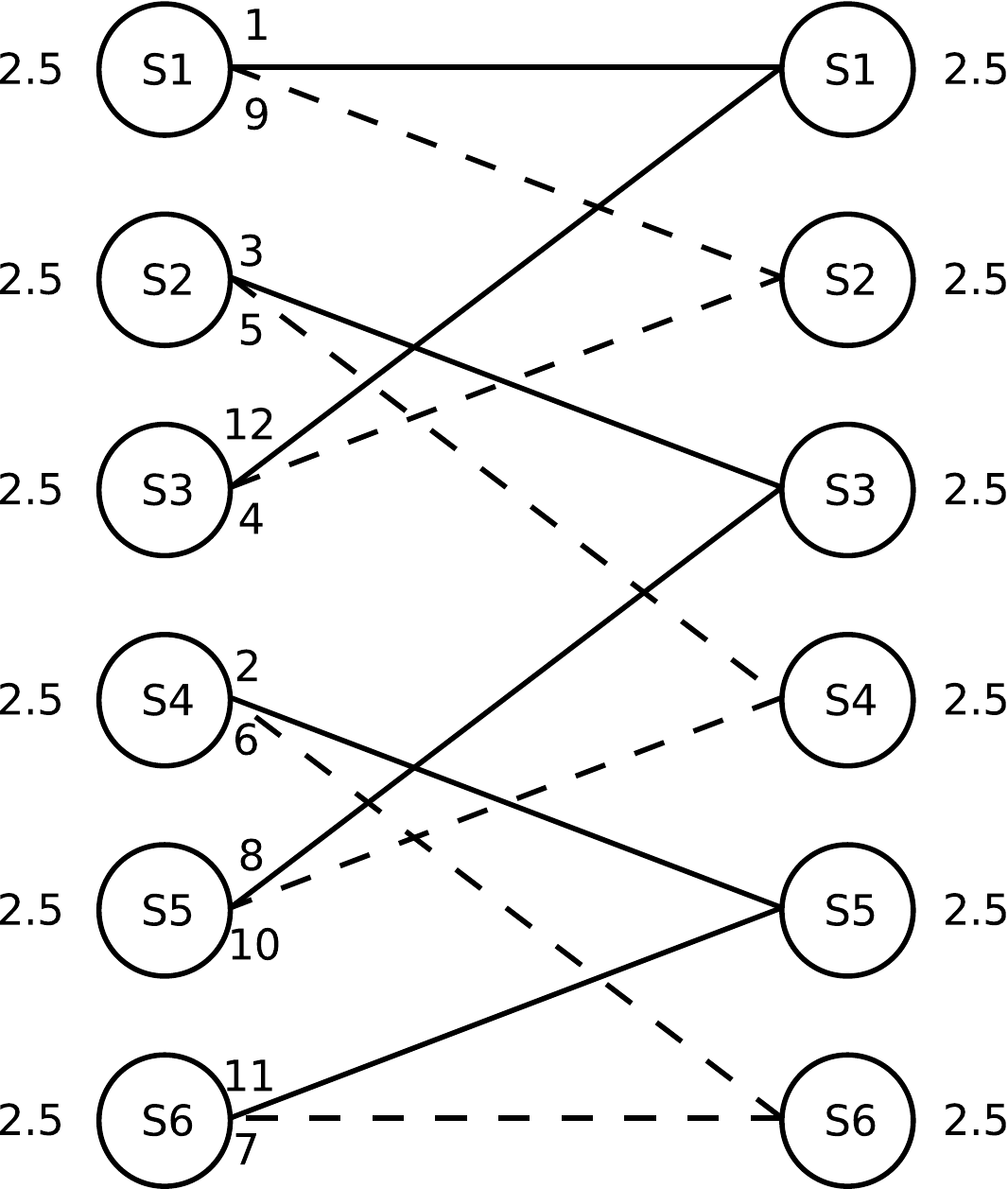}\\
{\footnotesize (c)} 
\end{minipage}
\caption{Code Design. a) Block diagram of the code. b) Trellis
construction for the set of multisymbols of Example 1. (c) Trellis
construction for the set of multisymbols of Example 2.}
\label{fig:Trellis}
\end{figure}

The coding scheme we propose is designed as a concatenation of an
outer error correcting code, an interleaver and an encoder, as
given in Fig.~\ref{fig:Trellis} a). The outer error correcting
code is a convolutional code with rate $1/2$, thus $2n$ binary
symbols are generated from $n$ symbols.
As already discussed, the inner code is trellis based, each branch
in the trellis is associated with an input symbol (binary) and
output symbol which is one of the $L$ multisymbols. We associate
multisymbols with the transitions in the trellis such that the
output symbols on the branches exiting from the same state should
be maximally separated in terms of expected Hamming distance.
The same is valid for the output symbols associated with the two
branches that enter the same state. The trellis encoder codes
incoming binary symbols into multisymbols which are then impaired
by the channel. In this part we assume a binary erasure channel.
However, we have in mind that the capacity results presented in
Part I are valid for a wider class of channels, among others the
binary symmetric channel and the $Z$ channel.

The symbol errors from a trellis code come in bursts since ending
up in a wrong state implies more than one symbol error. To avoid
bursts of errors, an interleaver is used. The interleaver is
implemented as matrix
with dimensions $\lambda \times \frac{2n}{\lambda}$ 
with $2n$ divisible by $\lambda$. In order to illustrate the
coding scheme, we take once again the example with $F=4$. The
trellis based coding scheme for $F = 4$ defines a trellis with
$12$ branches. One option is to consider a code with $4$ states
and $3$ branches from each state or a code, which implies that the
source information is originally encoded in ternary symbols.
Another, more practical option is to have a trellis with $6$
states and $2$ branches from each state. The code construction
uses the trellis code with $6$ states to avoid mapping from binary
symbols to ternary symbols. This means that one binary symbol is
transmitted for each multisymbol. The trellis design for the two
sets of multisymbols introduced in Section \ref{sec:CodingErrors}
with respective minimal distance $0.5$ and $1$ is presented on
Fig.~\ref{fig:Trellis}. We recall that the first set was described
by a directed graph where some of the multisymbols share common
edges which could be avoided. The trellis is given in
Fig.~\ref{fig:Trellis} b). The second set is chosen according to
the criterion which maximizes the minimal expected Hamming
distance. The trellis for the second set is illustrated in
Fig.~\ref{fig:Trellis} c). In both cases, the multisymbols which
are associated with the transitions in the trellis are chosen such
that the output symbols on the branches exiting from the same
state are maximally separated in terms of the expected Hamming
distance. Similarly, in both cases, the distance between the
output multisymbols is $2.5$ for all states. This is the maximal
distance in the distance spectrum. The same is valid for the
output symbols associated with the two branches that enter the
same state.

\subsection{Simulation Results} The simulations have been performed
with $\lambda = 8$ and the results are averaged over 10000
iterations. The simulation is performed for packet lengths
$N=\{2,6,14,30,62\}$. These packet lengths are chosen such that $N
+ 2$ (two tail bits are added by the outer convolutional code) is
divisible by $\lambda = 8$.

\begin{figure}
  \centering
  \includegraphics[width=9cm]{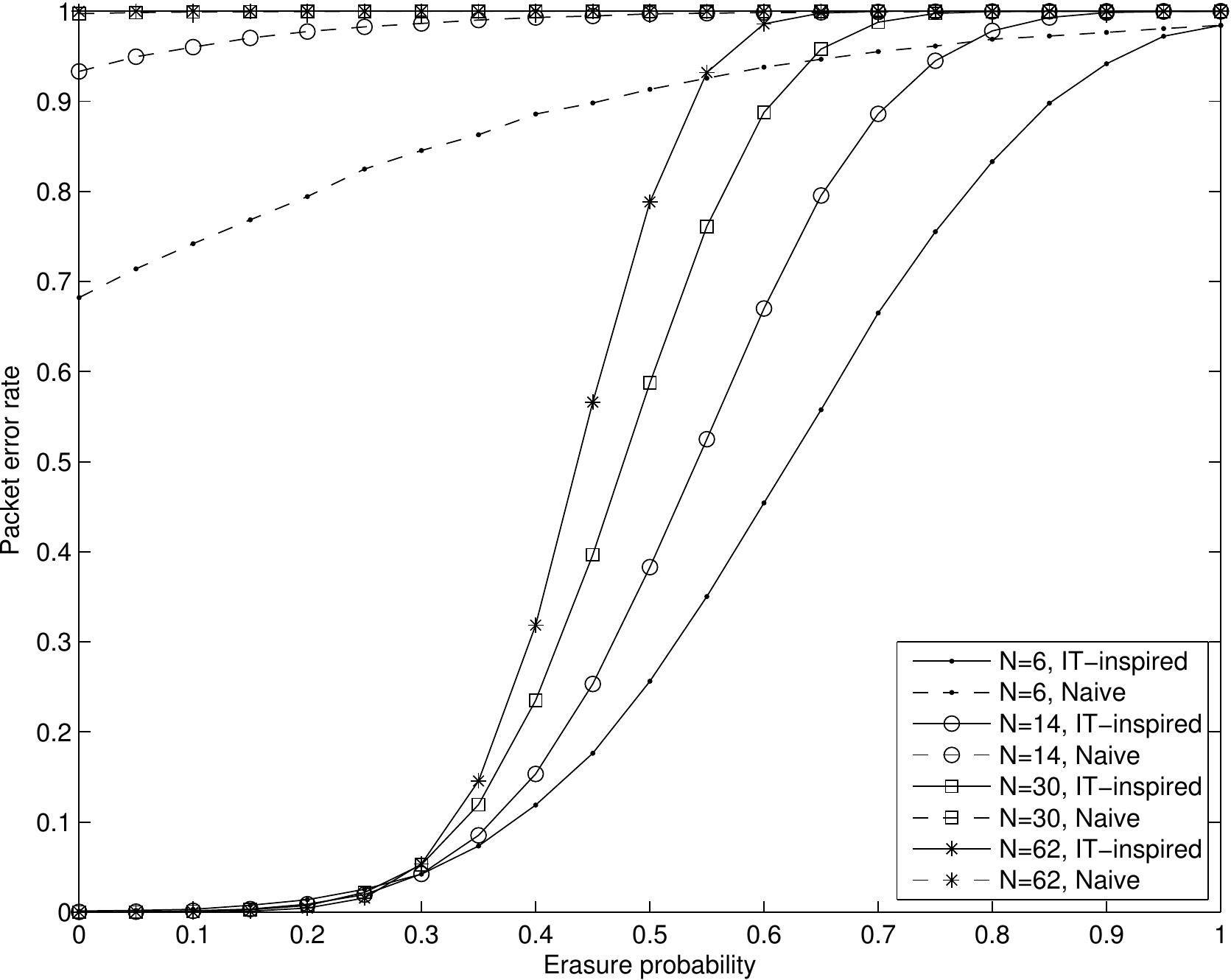}\\
  \caption{Performance comparison between the na\"{\i}ve coding scheme and the scheme motivated from the multiuser framework}\label{fig:Simulations1}
\end{figure}

First, we compare the performance of the coding scheme inspired by
the multisymbol framework and the na\"{\i}ve coding scheme, which
does not account for the specifics of the secondary channel. The
simulation results for the packet error rate (PER) for different
erasure probability are shown in Fig.~\ref{fig:Simulations1}. This
result present a clear evidence that the information-theoretic
analysis carries a practical significance for the secondary
communications channels.

\begin{figure}
\begin{minipage}[b]{0.5\linewidth}
\centering
\includegraphics[width=8cm]{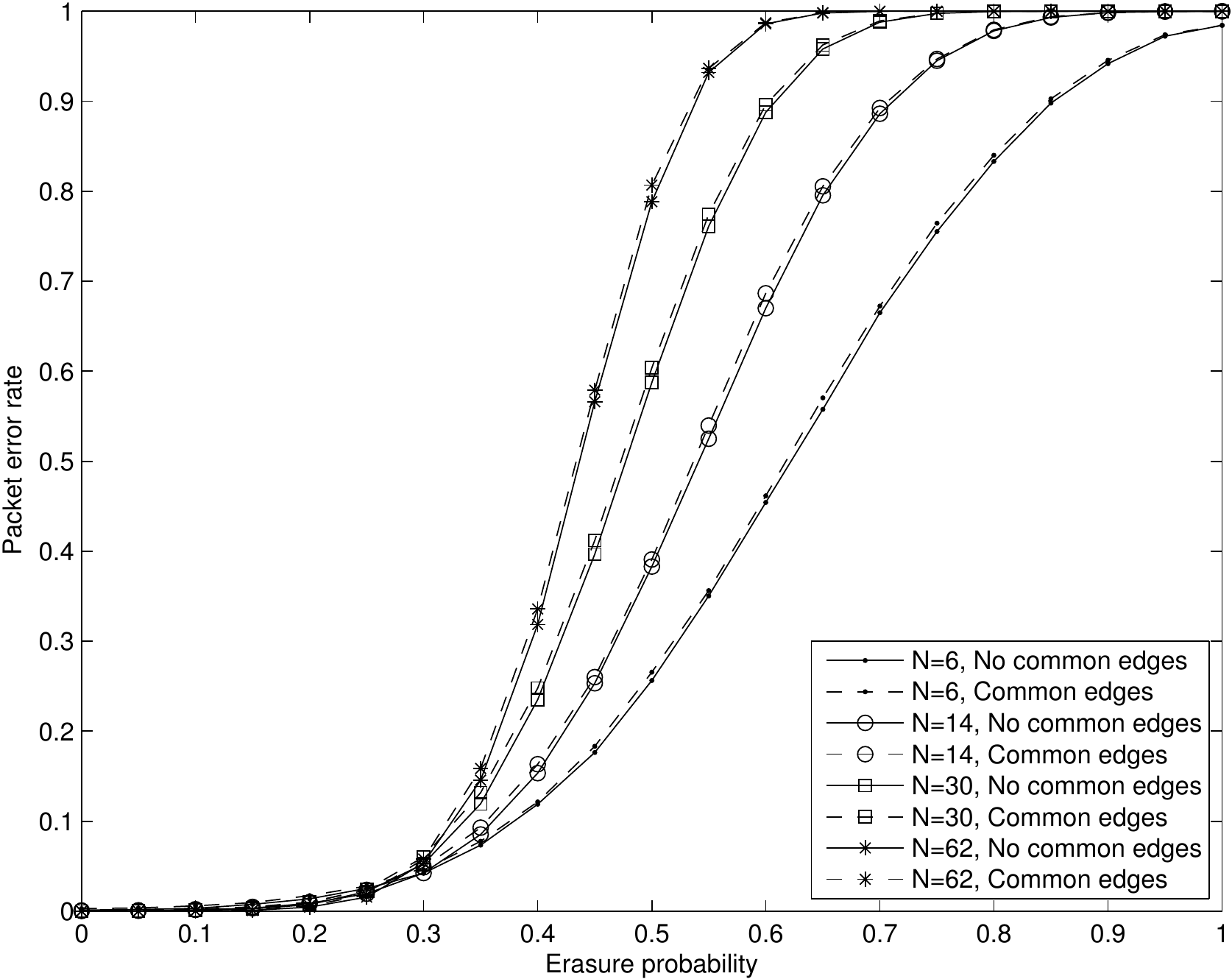}
{\footnotesize (a)} 
\end{minipage}
\hspace{0.5cm}
\begin{minipage}[b]{0.5\linewidth}
\centering
\includegraphics[width=8cm]{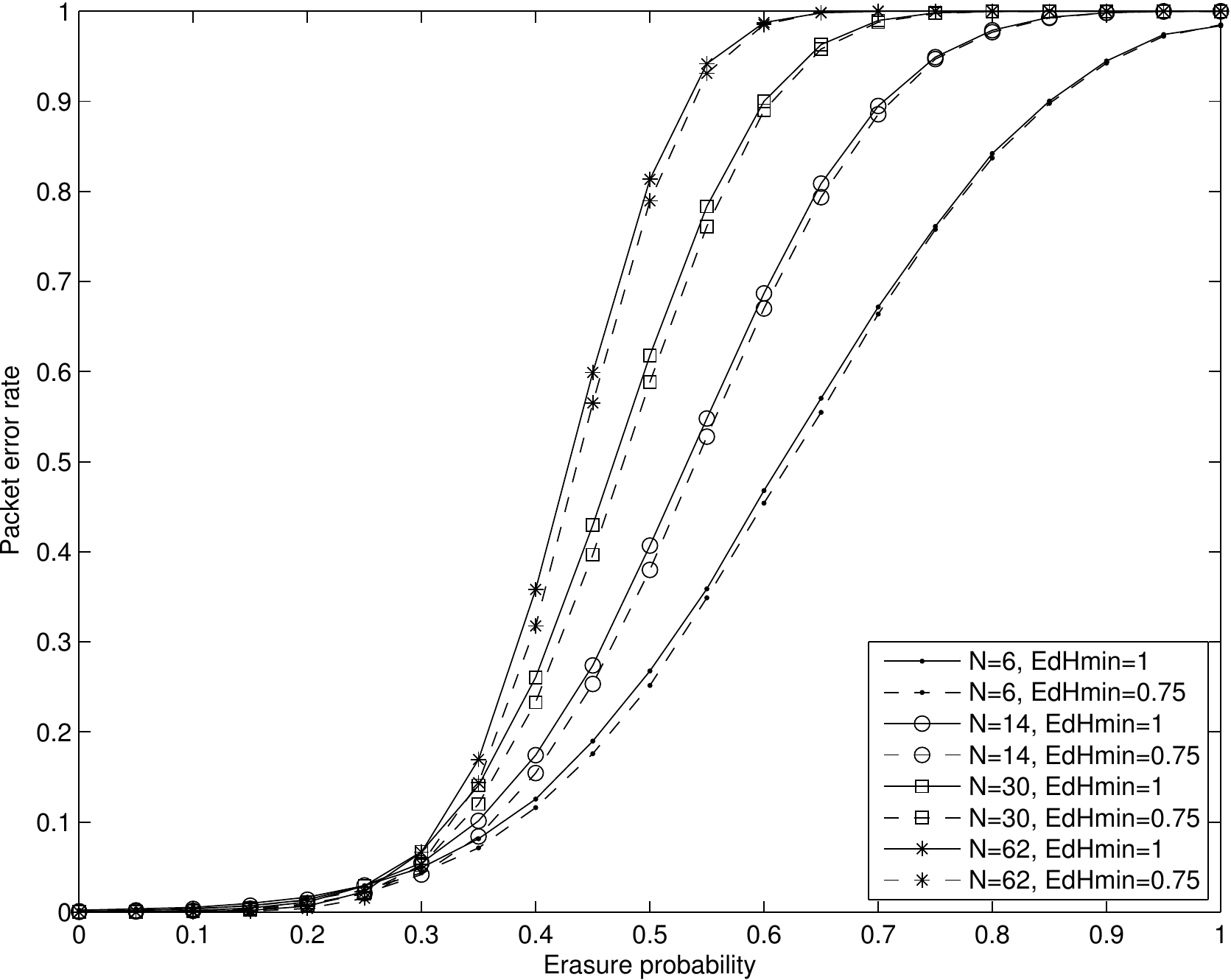}
{\footnotesize (b)} 
\end{minipage}
\caption{Performance of the error-correcting coding schemes.
  (a) Comparison between two sets with minimal distance $0.5$ and $1$
  respectively.
  (b) Comparison between two sets with minimal distance $0.75$ and $1$ respectively.}
  \label{fig:Simulations2}
\end{figure}

We also perform simulations for the two different choices of the
sets of multisymbols, with minimal expected Hamming distance $0.5$
and $1$ respectively, as presented in Section
\ref{sec:CodingErrors}. As already commented, although the choice
of the set with minimal distance $1$ performs better than the set
with minimal distance $0.5$ ( Fig.~\ref{fig:Simulations2} a)), we
were able to find another set with minimal distance $0.75$ which
outperforms both sets (Fig.~\ref{fig:Simulations2} a)). This
result indicates that besides the minimal expected Hamming
distance is not the only criterion which decides on the code
performance, but there are also factors,
notably the distance spectrum and the choice of the trellis transitions. 

\section{Discussion}
\label{sec:Discussion}

\subsection{Some Features of the Secondary Channel}

In Part I \cite{ref:Part1} we have described a generic application of this type of protocol coding and the associated secondary channel: communication with newly introduced devices, with limited functionality, in an area that is larger than the original coverage area.
Several distinctive features can be noted for this type of secondary communication. \emph{First}, for
delay-constrained systems, the secondary data rate is relatively
low, so it is hard to argue that this method brings a significant
rate advantage. \emph{Second}, the secondary rate depends on the
current load (traffic, number of users) in the primary system. For example, in a cellular system where protocol coding is done by encoding information in the way the users are allocated to different channels, the best secondary rate on is
obtained when each channel can be allocated to a different user, since this maximizes the number of possible rearrangements.
\emph{Third}, the new
secondary devices have a limited implementation of the primary protocol stack, which brings opportunity for a
low--complexity, low--power reception on the secondary channel. In
the extreme case, protocol coding reckons only with two
transmission states: packet transmitted and idle slot, as with the $Z-$channel model, which would require the secondary device to use only power detection.

Header compression~\cite{HeaderCompression} may appear as a
competitor as it works in a somewhat opposite way: tries to
compress the overhead whenever the actual communication scenario
allows it. However, this is not always cancelling the opportunity
for secondary communication, and vice versa: for example, the
MAC--layer identifiers may be compressed, but in the end all the
users have to be differentiated and the secondary channel arises
from reordering those identifiers. An interesting dividend is that
the secondary capacity can be used to assess the performance
margin of a certain primary protocol/system. Intuitively, if in a
given scenario the secondary capacity is non--zero, then the
operation of the primary system is not optimal.

The secondary channel, as defined here, can have several different application. A
generic application of the secondary communication is sending of
additional control data. The first usage of such a control data
can be as \emph{expanded ``future use'' bits}: in many
standardized protocols there are unspecified, free bits for future
use and protocol coding practically unleashes  ``hidden'' future
use bits in the protocol, which may become indispensable during
the evolution of the system.

Another usage can be signaling for efficient spectrum sharing. The
main concern in cognitive radio is the interference that the
cognitive (secondary) users are causing to the incumbent (primary)
user. Hence, a secondary user should sense if the spectrum
resource is available for communication. Spectrum sensing is
facilitated by a Cognitive Pilot Channel (CPC)~\cite{CPC}, which
conveys the necessary information to the terminals about the
status of radio spectrum. Protocol coding inherently introduces a
possibility to define an in--band CPC. For example, assume that, besides the module that can decode the
secondary channel, the secondary devices have an additional
cognitive radio interface to communicate with each other. Then the
primary BS can dynamically send information about the available
resources for cognitive radio. For example, if the primary system
is a digital TV broadcaster, then secondary channel can be defined
by reordering of the TV packets, which empowers the TV broadcast
tower to dynamically control the spectrum usage. To the best of
our knowledge, such a possibility to turn the TV broadcasters from
victims into spectrum controllers has not been observed before.

In the emerging machine--to--machine (M2M)
communication~\cite{ref:M2M}, cellular networks embrace a large number
of low--cost, low--power devices, that have different
traffic/behavior from the usual cellular users. Such a device
device is mostly in a low--power ``sleep'' mode. We conjecture
that, due to the simple codebooks used to send the primary control
information, it can be decoded with a low power. A sleeping device
may be tuned receive on the secondary channel and, upon receiving
a downlink trigger from the BS, it can wake up another radio
interface to send information. Thus, protocol coding offers an
opportunity to introduce universal wake--up beacons.

\subsection{Protocol Coding in WiMAX: A Brief Case Study}

We now illustrate the application of protocol coding with
reordering of user resources to the WiMAX system. In
WiMAX~\cite{WiMax_book}, the downlink and uplink control
information is transmitted at the beginning of each frame, which
includes preamble, frame control header (FCH) and MAP message. The
MAP message indicates the resource allocation for downlink and
uplink data and control signal transmission. The Base Station (BS)
translates the QoS requirements of the Subscriber Stations (SSs)
into the appropriate number of allocated slots. The BS informs
about the scheduling to all SSs by using the DL\_MAP (Downlink
Medium Access Protocol) and UL\_MAP (Uplink Medium Access
Protocol) messages in the beginning of each frame
\cite{Globecom_WiMax}.

Protocol coding is implemented by reordering the slots allocated
in a frame. The secondary users for which this information is
intended have only to read the broadcast DL\_MAP and UL\_MAP
messages. For example, when the number of slots reserved for each
of the SSs is 6,9,2,10,7,6,10,15,15,20 respectively, $289$
secondary bits can be sent by reordering of the resources.
Assuming a frame duration of $5 ms$, this translates to we can
have $\approx 58$ [kbps] of additional information, which is in
the frame headers that are robustly protected~\cite{Wang}. In
order to get an idea about the the distance where the MAP message
is ``detectable'', compared to the information data, we resort to
the propagation model in \cite{Wang}, with the total path loss is
given by $L=126,2+36 \log d$ [dB], where $d$ is in kilometers. The MAP is
protected with $6-$times repetition coding, while and BPSK is used
for both MAP message and data, which results in distance $d'
\approx 1.65 \: d$ where the header is detectable compared to the
distance $d$ for the user data.

\subsection{Further Considerations}
\label{sec:FurtherConsiderations}

We used a simplified model, in which the set of packets sent in a
given frame is independent from the other frames. In practice this
is rarely satisfied, since buffering at the primary scheduler
and/or packet retransmission due to errors creates dependencies
between consecutive frames. In such a case, Shannon's result is
not directly applicable and instead we need to use a more general
model in which the sequence of frame states is not memoryless (see
Section 6 in~\cite{ref:CSITmonograph}).

Another aspect is the freedom of in reordering user resources. For
example, if in the case of WiMAX the scheduler puts each user on a
channel where she can achieve a high data rate, then the freedom
to permute users across channels becomes restricted. It is
incorrect to say that protocol coding is not applicable once such
restrictions are put by the primary system, but it should rather
be observed that the secondary capacity is decreased. This
reiterates the observation that protocol coding can be used as a
measure of how optimally given primary system operates.

We have mainly discussed the case of the combinatorial model with
two possible packet values $0,1$. In general, the number $m$ of
possible packets in a frame can be $1 \leq m \leq F$, where the
special case $m=F$ corresponds to the permutation model. As coding
strategies, it is relevant to consider the permutation codes used
in power line communication(PLC)~\cite{HanVinck}. The main idea is
to send information by using $M-$ary Frequency Shift Keying (FSK).
There are $M$ available orthogonal frequencies and transmission is
done by creating a time sequence by which the frequencies are
activated. This corresponds to a permutation without repetitions
of size $M$ and the correspondent codes can be used as coding
strategies in our permutation model. The main idea is to define a
Hamming distance between two permutations and only permutations
that are sufficiently distant are eligible for transmission in
order to minimize the probability of error. A more general case is
the one where the transmission symbol (frame) has $F>M$
frequencies and the $i-$th frequency appears $f_i$ times, such
that $\sum_{i=1}^M f_i=F$. The codes used in that case are termed
\emph{constant composition codes}~\cite{PermCodesDesign}. However,
the main design constraint for permutation codes in PLC scenarios
is not to create a disturbance to the electric power. This implies
that there is a freedom to choose the set $ \{ f_i \}$ as long as
the power constraint is satisfied; on the contrary, in our model
the secondary transmitter must reckon with the set $ \{ f_i \}$
provided by the primary communication system. Nevertheless, the
code design from frequency permutation arrays in PLC may be used
to select the components of the multisymbols for protocol coding.

\section{Conclusion and Future Work}
\label{sec:Conclusion}
We have investigated practical
strategies for protocol coding via combinatorial
ordering of the user resources (packets, channels) in the primary
system. By using the specific structure of our model, we have used
the alternative framework that helps to compute the capacity in
the development of coding strategies that are approaching the
capacity. The developed coding strategies are superior to the
na\"{\i}ve strategy which does not account for the specifics of the
secondary communication channels. The coding design thus gives
practical relevance to the framework developed for capacity
characterization of secondary channels and paves the road for
practical implementation. Besides the construction of the coding
schemes, we also presented some additional features of protocol
coding and pointed out possible applications of the concept in
existing wireless systems.

A question for future work is how to compute the capacity and
which coding strategies to use when the scheduling process in the
primary system is generalized (buffering, retransmission, etc.).
Another direction is to compute the capacity under error models
different from the described, such as channels with
deletions/insertions. In practice, a secondary channel can be
defined over virtually any existing wireless system and therefore
it is of interest to find the coding strategies that are suited to
the actual protocol specification in a certain primary system.

\bibliographystyle{IEEEtran}
\bibliography{references}

\end{document}